\documentclass[conference]{IEEEtran}
\IEEEoverridecommandlockouts

\usepackage{cite}
\usepackage{bbm}
\usepackage{amsmath}
\usepackage{amssymb}
\usepackage{outlines}
\usepackage{array}
\usepackage{float}
\usepackage{kotex}
\usepackage{graphicx}
\usepackage{bm}
\usepackage{caption,subcaption}
\usepackage{algorithm}
\usepackage{algpseudocode}
\usepackage{csquotes}
\usepackage{cite}
\usepackage{multicol}
\usepackage{stmaryrd}
\usepackage{tabularx,booktabs}
\usepackage{xcolor}
\usepackage[
            top=0.75in,
            bottom=1in,
            left=0.625in,
            right=0.625in]{geometry}
\newcolumntype{C}{>{\centering\arraybackslash}X} 
\usepackage{lipsum} 

\def\BibTeX{{\rm B\kern-.05em{\sc i\kern-.025em b}\kern-.08em
    T\kern-.1667em\lower.7ex\hbox{E}\kern-.125emX}}
    
\begin{document}
\title{Enhancing Battlefield Awareness: 
An Aerial RIS-assisted ISAC System with Deep Reinforcement Learning}
\author{\IEEEauthorblockN{\textrm{Hyunsang Cho}$^*$, \textrm{Seonghoon Yoo}$^*$, \textrm{Bang Chul Jung}$^\dagger$, \textrm{Joonhyuk Kang}$^*$}
\IEEEauthorblockA{$^*$\textrm{Korea Advanced Institute of Science and Technology (KAIST), Daejeon, Korea} \\
$^\dagger$\textrm{Chungnam National University, Daejeon, Korea}\\
Email: $^*$ \textrm{\{whgustkdi, shyoo902\}@kaist.ac.kr }$^\dagger$ \textrm{bcjung@cnu.ac.kr,}  $^*$ \textrm{jhkang@kaist.edu} }

\thanks{{This work was supported in part by the Ministry of
Science and ICT (MSIT), South Korea, through the Information Technology Research Center (ITRC) Support Program supervised by the Institute of Information and Communications Technology Planning and Evaluation (IITP)
under Grant IITP-2024-2020-0-01787 and Grant IITP-2024-RS-2023-00259991.}}
\thanks{Hyunsang Cho and Seonghoon Yoo contributed equally to this work.}
}
\maketitle

\begin{abstract}
\textcolor{black}{This paper considers a joint communication and sensing technique for enhancing situational awareness in practical battlefield scenarios. In particular, we propose an aerial reconfigurable intelligent surface (ARIS)-assisted integrated sensing and communication (ISAC) system consisting of a single access point (AP), an ARIS, multiple users, and a sensing target. With deep reinforcement learning (DRL), we jointly optimize the transmit beamforming of the AP, the RIS phase shifts, and the trajectory of the ARIS under signal-to-interference-noise ratio (SINR) constraints. Numerical results demonstrate that the proposed technique outperforms the conventional benchmark schemes by suppressing the self-interference and clutter echo signals or optimizing the RIS phase shifts.}
\end{abstract}

\section{Introduction}
The demand for advanced wireless sensing and communication technologies is increasing as the battlefield environment becomes more complex and dynamic due to the increase in the variety of devices. Recently, integrated sensing and communication (ISAC) has been recognized as a promising technology for future wireless networks using high-frequency bands, e.g., millimeter wave (mmWave) \cite{isac_overview}. In particular, ISAC can potentially enhance overall operational efficiency on the battlefield, given that radar sensing and wireless communication share the same spectrum and hardware facilities \cite{military}.    

The overall process of the ISAC downlink system generally proceeds with the access point (AP) transmitting ISAC signals to users and processing echo signals reflected from the target. However, due to the dominant line-of-sight (LoS) channel characteristics of the links, ISAC in military scenarios cannot avoid the problem of being blocked by various obstacles, e.g., mountains, and causing severe path loss as the communication distance increases \cite{clutter_user}. To overcome the physical limitations of the LoS channel, reconfigurable intelligent surface (RIS) is emerging as a key technology that expands target detection and communication range by adjusting the phase shifts to reconfigure signal propagation \cite{ris, ruiizhang_RIS_ISAC}. The authors in \cite{ruiizhang_RIS_ISAC} propose the joint transmit and receive beamforming in a RIS-aided single-target multi-user ISAC system. Deploying a terrestrial RIS between an AP and a ground node, however, has limitations in providing sufficient quality-of-service (QoS) in the dynamic battlefield environment. On the other hand, aerial-RIS (ARIS), which mounts RIS on an unmanned aerial vehicle (UAV), can provide more effective sensing and communication performance in a dynamic battlefield environment by leveraging mobility \cite{ARIS}. In \cite{ARIS_ISAC}, the ISAC system assisted by ARIS is considered to reconfigure the propagation environment to counter malicious jammers flexibly.

The solutions in previous works \cite{ARIS,ARIS_ISAC} addressing ARIS systems for sensing or communication networks are mostly provided by convex optimization and cannot be quickly applied to battlefield scenarios. The deep reinforcement learning (DRL) approach has been actively adopted as an alternative to conventional optimization approaches due to its advantage of deriving policies while interacting with the environment through a deep neural network. Among DRL algorithms, deep deterministic policy gradient (DDPG) is known to converge and operate well in continuous action spaces such as ARIS trajectories \cite{ddpg}. The authors of \cite{benchmark} propose a DRL-based ARIS trajectory design for communication and localization with a vehicle. However, from a practical perspective, the self-interference problem \cite{self_interference_not_neglect} cannot be ignored when the AP operating in full-duplex mode, and a method to suppress clutter echo signals \cite{clutter_user} is also required.

This work focuses on the DRL-based ARIS-assisted ISAC system in military scenarios, where a multi-antenna AP serves ground users and detects a target. We aim to minimize the Cramér-Rao bound (CRB) \cite{crb} for target localization by jointly optimizing the transmit beamforming, the RIS phase shifts, and the ARIS trajectory. Moreover, to address the challenge posed by self-interference and clutter echo signals, we leverage a null-space projection (NSP)-based receive beamforming scheme \cite{nsp} that suppresses these signals. To cope with the non-convexity of the formulated problem, we propose a DDPG-based algorithm that searches for the optimal policy while interacting with the environment. Via simulations, the proposed method is verified to outperform other benchmark methods, such as fixing the RIS phase shifts or not applying the NSP-based receive beamforming scheme.

The rest of the paper is organized as follows: In Section II, we introduce the system model, including the channel, communication, and radar sensing models of the ARIS-assisted ISAC systems. Section III presents the proposed DRL-based algorithm that aims to minimize the overall system's CRB. Numerical results are shown in Section IV, and Section V concludes the paper.

\textbf{Notation:} In this paper, we denote $(\cdot)^H$ as complex conjugate transpose; $(\cdot)^*$ as complex conjugate;  $(\cdot)^T$ as transpose; $\lVert \cdot \rVert$ denotes the Euclidean norm; $\odot$ denotes the Hadamard (element-wise) products; $\succeq$ as positive semi-definite; and diag$(a_1,\cdots,a_N)$ denotes a diagonal matrix with diagonal elements $a_1,\cdots, a_N$.\\

\section{System model}
\begin{figure}[t] 
\begin{center}
\includegraphics[width=1\columnwidth]{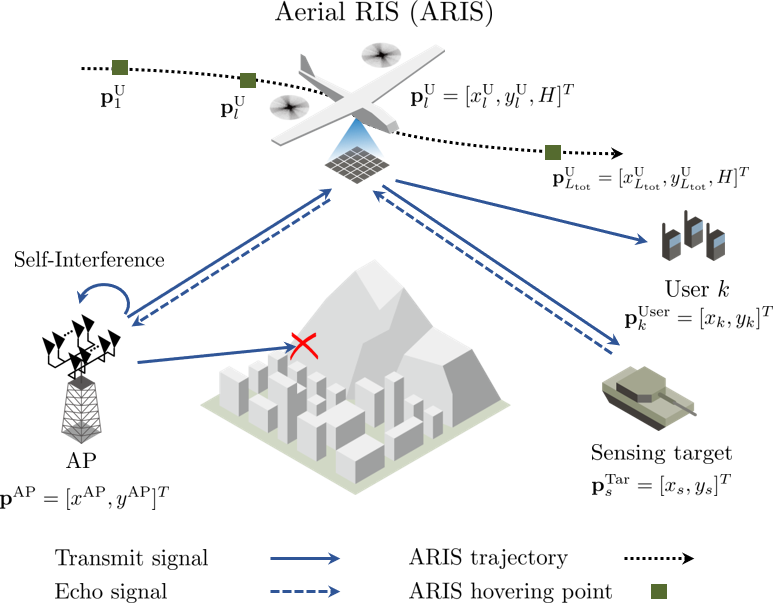}
\caption{Illustration of the proposed ARIS-assisted ISAC systems.}
\label{fig1}
\end{center}
\end{figure}
\subsection{Set-up}
We consider a battlefield condition, which assumes that the direct LoS links between the AP and nodes in the battlefield are blocked, as depicted in Fig. \ref{fig1}. In these situations, we propose an ARIS-ISAC system, which consists of a full-duplex AP with a $M$ uniform linear array (ULA) antennas, an ARIS with $N$ passive transmission-reflection units to serve $K$ single-antenna ground users, and a point sensing target. An ARIS is deployed to create virtual LoS links leveraging mobility to support communications and facilitate target detection \cite{ARIS}. In the scenario we cover, we assume a point-sensing target consisting of a single path because the distances between links are long in a far-field environment, and the spatial extent is relatively small \cite{RIS_ISAC}. For traceability of analysis, ARIS operates for a total $L_{\rm{tot}}$ time slots (TSs), and each slot becomes a hovering point for ARIS to perform both sensing and communication, assuming that the channel does not change for $\Delta$ seconds \cite{delta_seconds}. In the following, we use the notations of `$\text{U}$' for the ARIS. ARIS hovers at a fixed altitude $H$ and flies at the velocity variable $\mathbf{v}^\mathrm{U}_l=\{v_{x,l}, v_{y,l}\}$ with the horizontal velocity 
$v_{x,l}$ and the vertical velocity $v_{y,l}$ in TS $l\in \{1,2,...,L_{\rm{tot}}\}\triangleq \mathcal{L}$, both of which are limited by the maximum velocity constraint $v_{\max}$. Accordingly, ARIS's coordinates can be expressed as $\mathbf{p}^\text{U}_l=(x^\text{U}_l,y^\text{U}_l,H)^T$, where $x^\text{U}_l=x^\text{U}_0+\sum_{l'=1}^{l}v_{x,l'}\Delta$ and $y^\text{U}_l=y^\text{U}_0+\sum_{l'=1}^{l}v_{y,l'}\Delta$. The AP, sensing target and the user $k \in \mathcal{K}\triangleq\{1,2,...,K\}$ are on the ground and assumed to be fixed at $\mathbf{p}^{\text{AP}}=(x^{\text{AP}},y^{\text{AP}},0)^T$, $\mathbf{p}_s^{\text{Tar}}=(x_s^{\text{Tar}},y_s^{\text{Tar}},0)^T$ and $\mathbf{p}_k^{\text{User}}=(x^{\text{User}}_{k},y^{\text{User}}_{ k},0)^T$, respectively.

To demonstrate the ISAC signal transmitted
at the AP at the TS $l$, let $s_{k,l}$ denotes the transmit communication signal by the AP at TS $l \in \mathcal{L}$, and $\mathbf{w}_{k,l} \in \mathbb{C}^{M \times 1}$ denotes the corresponding transmit beamforming vector. The dedicated sensing signal vector at time slot $l$ is given by $\mathbf{x}_{s,l}= \mathbf{w}_{s,l}\odot \mathbf{s}_{s,l} \in \mathbb{C}^{M \times 1}$, where $\mathbf{w}_{s,l}$ and $\mathbf{s}_{s,l}$ are the corresponding radar beamforming vector and transmit signals, respectively. By following \cite{ruiizhang_RIS_ISAC}, the sample covariance matrix of $\mathbf{x}_{s,l}$ is $\mathbf{R}_{s,l}=\mathbb{E}[\mathbf{x}_{s,l}\mathbf{x}_{s,l}^H] \succeq 0 , \forall l \in \mathcal{L}$. 
Then, the transmitted signal $\mathbf{x}_l\in \mathbb{C}^{M \times 1}$ by the AP at TS $l$ is expressed as
\begin{equation}
    \mathbf{x}_l=\sum_{k \in \mathcal{K}}\mathbf{w}_{k,l} s_{k,l}+\mathbf{x}_{s,l}.                    
\end{equation}
Here, we consider a total transmit power constraint as $\sum\nolimits_{k\in \mathcal{K}}\Vert \mathbf{w}_{k,l}\Vert^{2}+\mathrm{t}\mathrm{r}\left[\mathbf{R}_{s,l}\right]\leq P_{\text{AP}}, \forall l \in \mathcal{L}$, where $P_{\text{AP}}$ denotes the maximum power budget of the AP.

In particular, the RIS shifts the phases of the reflecting elements with unit amplitudes. RIS not only provides a communication link between the AP and the user, but also allows the AP to receive echo signals for target detection \cite{RIS_ISAC}. The reflection matrix in TS $l$ can be defined as $\mathbf{\Phi}_l=\mathrm{diag}(e^{j\theta_1},...,e^{j\theta_N}) \in \mathbb{C}^{N\times N}$, where $\theta_n$ denotes the phase-shift of the $n$-th element of the RIS. 

\subsection{Channel model}
At each TS $l$, all channel coefficients are assumed to be quasi-static flat fading. This implies that the channel power gain $\mathbf{G}^{\text{AP,U}}_l \in \mathbb{C}^{N\times M}$ between AP's $m$-th antenna element and $n$-th RIS element of ARIS in TS $l$ \cite{uav_channel} can be written as
\begin{equation}
  [\mathbf{G}^{\text{AP,U}}_l]_{n,m}=\frac{\beta_0}{\lVert \mathbf{p}^{\text{AP}}-\mathbf{p}^{\text{U}}_l\rVert ^2}, \forall n \in \mathcal{N}, \forall m \in \mathcal{M},
\end{equation}
where $\beta_0$ denotes the reference channel power gain of the G2A link. Similarly, for the channel power gain between ARIS and user $k$, we define $\mathbf{h}_{l}^{\text{U},k} \in \mathbb{C}^{N\times 1}$ as
\begin{equation}
    [\mathbf{h}_{l}^{\text{U},k}]_{n}=\frac{\beta_0}{\lVert \mathbf{p}^{\text{U}}_l-\mathbf{p}^{\text{User}}_k\rVert ^2}, \forall k \in \mathcal{K}, \forall n \in \mathcal{N}.
\end{equation}
We consider a LoS model for the ARIS-target link to facilitate departure-of-arrival (DoA) estimation. 
Let $\theta_i$ denote the DoA with respect to the RIS, and the steering vector at the RIS with angle $\theta_i$ is given by $\mathbf a_l(\theta _{i})=[1,e^{j2\pi d_{\text{RIS}}\sin\theta/\lambda},...,e^{j2\pi(N-1)d_{\text{RIS}}\sin\theta/\lambda}]^T \in \mathbb{C}^{N\times 1}$, where $d_{\text{RIS}}$ denotes the spacing between elements at the RIS and $\lambda$ denotes the wavelength. The $i$ element must consider the target and the clutters, i.e., users, during the reflection process \cite{clutter_user}. In this case, the target response matrix $\mathbf{H}_l \in \mathbb{C}^{N\times N}$ for the ARIS-target-ARIS link is modeled as the summation of target reflection $\mathbf{H}^{\text{Tar}}_l $ and echo signal from the clutters $\mathbf{H}^{\text{C}}_l$ as follows:
\begin{equation}
\mathbf{H}_l=\mathbf{H}^{\text{Tar}}_l + \mathbf{H}^{\text{C}}_l = \underbrace{ \alpha_{s,l} \mathbf{A}_l(\theta_{s})  }_{\text{Target reflection}} + \underbrace{ \sum\nolimits_{k=1}^{K} \alpha_{k,l} \mathbf{A}_l(\theta_{k})}_{\text{Echo signal of clutters}},
\label{target_response}
\end{equation}
where $\mathbf A_l(\theta_{i}) \triangleq \mathbf a_{l}(\theta_{i})\mathbf a_{l}^H(\theta _{i})$ . Additionally, the two-way channel power gain between the ARIS and node $i \in \{s,1,...,K\}$ in TS $l$ is denoted by $\alpha_{i,l} = \sqrt{{\beta_{s}}/{\left(d_{i,l}\right)^4}}$), where $d_{i,l}$ represents the distance from ARIS to the node $i$ and $\beta_{s}$ is the reference channel power gain considering radar cross-section (RCS) over target when $d_{i,l} = 1 \textrm{m}$. 

Without loss of generality, a full-duplex AP cannot avoid additional signal attenuation problems that occur through self-interference (SI) in receiving sensing echo \cite{self_interference_not_neglect}. According to \cite{clutter_user, self-interference}, residual SI channel can be expressed as $[\mathbf G^{\text {SI}}]_{i,j} = \sqrt {\gamma ^{\text {SI}}_{i,j}} e^{-j2\pi {d_{i,j}}/{\lambda }}, \forall i,j \in \mathcal{M}$, where $\gamma ^{\text {SI}}_{i,j}$ and $d_{i,j}$ denote the residual SI channel power and the distance between the $i$-th transmit antenna and $j$-th receive antenna.

\subsection{Communication Model}
This section considers wireless communication signals transmitted from the AP and received by user $k$ via ARIS. The received signal by user $k$ at TS $l$ is
\begin{equation}
 \begin{aligned}
y_{k,l}&= (\mathbf{h}^{\text{U},k}_l)^H \mathbf{\Phi}_{l} \mathbf{G}^{\text{AP,U}}_l \mathbf{x}_l+n_{k,l} \\
&= \underbrace{ (\mathbf{h}^{\text{U},k}_l)^H \mathbf{\Phi}_l \mathbf{G}^{\text{AP,U}}_l \mathbf{w}_{k,l} s_k}_{\text{desired information signal}}+\hspace{-0.07cm} \underbrace{(\mathbf{h}^{\text{U}, k}_l)^H \boldsymbol{\Phi}_l \mathbf{G}^{\text{AP,U}}_l\hspace{-0.2cm}\sum_{i \in \mathcal{K}, i \neq k}\hspace{-0.2cm} \mathbf{w}_{i,l} s_i}_{\text{inter-user interference}} \\
&\quad+ \underbrace{(\mathbf{h}^{\text{U},k}_l)^H \boldsymbol{\Phi}_l \mathbf{G}^{\text{AP,U}}_l \mathbf{x}_{s,l}}_{\text{sensing signal interference}} +n_{k,l}, \,\,\,\,\forall k \in \mathcal{K}, \forall l \in \mathcal{L},
\end{aligned}   
\end{equation}
where $n_{k,l} \sim \mathcal{CN}(0,\sigma_k^2)$ denotes additive white Gaussian noise (AWGN) at the receiver of user $k$ with noise power $\sigma_k^2$. Considering sensing signal interference and inter-user interference \cite{ruiizhang_RIS_ISAC}, the corresponding signal-to-interference-noise ratio (SINR) of user $k$ can be denoted as
\begin{equation}
\Gamma_{k,l}= \frac{\left\vert \mathbf{h}_{k,l}^H \mathbf{w}_{k,l}\right\vert ^2}{\sum_{i \in \mathcal{K}, i \neq k}\left\vert \mathbf{h}_{k,l}^H \mathbf{w}_{i,l}\right\vert ^2+\mathbf{h}_{k,l}^H \mathbf{R}_{s,l} \mathbf{h}_{k,l}+\sigma_k^2}
\end{equation}
, $\forall k \in \mathcal{K}, \forall l \in \mathcal{L}$,
where $\mathbf{h}_{k,l} \triangleq (\mathbf{G}^{\text{AP,U}}_l)^{H} \boldsymbol{\Phi}_l^{H} \mathbf{h}^{\text{U},k}_l$.
\subsection{Radar Sensing Model}
We first implement an effective receive beamforming method for the radar sensing model that suppresses both SI and clutter echo. Then, the CRB of target coordinates is derived as the sensing performance.
\subsubsection{Receive beamforming design}
The AP needs to design effective receive beamforming that can properly receive the target echo from signals generated by other components. The received echo signal by the AP receiver $\mathbf{z}_{s,l} \in \mathbb{C}^{M\times 1}$ in TS $l$ can be written as
\begin{equation}
\begin{aligned}
\mathbf{z}_{s,l}=&\big((\mathbf{G}_l^{\text{AP,U}})^T\mathbf{\Phi}_l^T\mathbf{H}^{\text{Tar}}_l\mathbf{\Phi}_l\mathbf{G}_l^{\text{AP,U}}\\&+\underbrace{(\mathbf{G}_l^{\text{AP,U}})^T\mathbf{\Phi}_l^T\mathbf{H}^{\text{C}}_l\mathbf{\Phi}_l\mathbf{G}_l^{\text{AP,U}}}_{\text{Clutter echo}}+\underbrace{\mathbf{G}^{\text{SI}}}_{\text{SI}}\big)\mathbf{x}_{l}+\mathbf{n}_{s,l},
\end{aligned}
\end{equation}
where $\mathbf{n}_{s,l} \sim \mathcal{C N}(\mathbf{0}, \sigma^2_{s}\mathbf I_M)$ is the noise generated by each antenna of the ULA in the AP and $\sigma^2_{s}$ is the noise power at the AP. To effectively suppress echoes from clutters in equation (\ref{target_response}) and SI, we propose the null-space projection (NSP) method \cite{nsp}. After projecting the signal in the null-space direction of the SI and the clutter echo signals, the receive beamforming matrix $\mathbf{f}_{s,l} \in \mathbb{C}^{M \times 1}$ is derived in a closed form as follows:
\begin{equation}
    \mathbf{f}_{s,l} = \frac{{{{\big( {\left( {{\mathbf{I}} - {\mathbf{C}}{{\mathbf{C}}^ + }} \right)(\mathbf{G}_l^{\text{AP,U}})^T\mathbf{\Phi}_l^T{{\mathbf{a}}_{l}}\left( {{\theta _{s}}} \right)} \big)}}}^*}{{\left\| {\left( {{\mathbf{I}} - {\mathbf{C}}{{\mathbf{C}}^ + }} \right)(\mathbf{G}_l^{\text{AP,U}})^T\mathbf{\Phi}_l^T{{\mathbf{a}}_{l}}\left( {{\theta _{s}}} \right)} \right\|}}, \quad \forall l \in \mathcal{L},
\end{equation}
where $\mathbf{C}\hspace{-0.05cm}=\hspace{-0.05cm}[\mathbf{G}^{\text{SI}}\mathbf{w}_{1,l},... ,\mathbf{G}^{\text{SI}}\mathbf{w}_{K,l}, \mathbf{G}^{\text{SI}}\mathbf{w}_{s,l}, (\mathbf{G}^{\text{AP,U}}_l)^T\mathbf{\Phi}_l^T\mathbf{a}_{l}\left( \theta_{1} \right)\\,..., (\mathbf{G}^{\text{AP,U}}_l)^T\mathbf{\Phi}_l^T\mathbf{a}_{l}\left( \theta_{K} \right)]$. Finally, we then derive the received signal $\hat{z}_{s,l}={\mathbf{f}}_{s,l}^T\mathbf{z}_{s,l}$ that can more precisely estimate the target coordinates.
\subsubsection{CRB Derivation of Target Coordinates}
Our goal is to estimate the target coordinates accurately in the battlefield situation. CRB can provide a lower bound for the minimum-squared error (MSE) of any unbiased estimator \cite{crb}. Therefore, we aim to minimize the CRB, which is the inverse matrix of the Fisher information matrix (FIM) for estimating the target coordinates ${\mathbf{p}_s^{\text{Tar}}}=(x_s,y_s)$ as the sensing performance metric. In Chapter II-D, we can estimate the distance between the target and ARIS ${\mathbf{d}}_s=\left[d_{s,1},...,d_{s,L_{\rm{tot}}}\right]^T$ based on the estimated target DoA for each TS through the received target echo signal. Therefore, the FIM of the distance between ARIS and the target $\mathbf{J}(\mathbf{d}_{s})$ is first constructed, and then the FIM of the target coordinates $\mathbf{J}(\mathbf{p}_{s}^{\text{Tar}})$ is derived through mathematical relationships, which is adopted in recent works \cite{isac_fromthesky, uav_ISAC_crb}.

According to \cite{snr_inverse_proportional}, the variance of measurements $\widehat{\mathbf{d}}_s$ is inversely proportional to the received signal-to-noise ratio (SNR) of the ISAC echo as
\begin{equation}
\label{variance}
    [\sigma({d}_{s,l})]^2= \frac{a}{\mathrm{SNR}_{s,l}} = \frac{a  \sigma^2_{s}}{P_{\text{AP}}  G_{{p}}\lVert(\mathbf{G}_l^{\text{AP,U}})^T\mathbf{\Phi}_{l}^T\mathbf{H}^{\text{Tar}}_l\mathbf{\Phi}_l\mathbf{G}_l^{\text{AP,U}}\rVert^2},
\end{equation}
where $a$ is a pre-determined constant considering various factors, i.e., terms for communication between ARIS and AP, and $G_p$ is the signal processing gain. Therefore, measurement covariance $\mathbf{C}({\mathbf{d}}_s)$ of the distribution that the measurement vector follows with ${\mathbf{\widehat{d}}_s} \sim \mathcal{N}(\mathbf{d}_s,\mathbf{C}(\mathbf{d}_s))$ can be obtained through a total of $L_{\rm{tot}}$ hovering points. With this assumption, we can derive the general expression of $p, q$-th element in $\mathbf{J}(\mathbf{d}_s)$ according to \cite{measurementcovariance}. The detailed process of deriving the $\mathbf{C}({\mathbf{d}}_s)$ and $\mathbf{J}(\mathbf{d}_s)$ are introduced in the Appendix.

Based on the FIM for the distance, we can obtain the FIM of target coordinates as ${\mathbf{J}}\left({\mathbf{p}_s^{\text{Tar}}}\right)={\mathbf{Q}}{\mathbf{J}}\left({\mathbf{d}}_{s}\right){\mathbf{Q}}^T$, where ${\mathbf{Q}}={\partial{(\mathbf{d}}_s)^T}/{\partial{\mathbf{p}_{s}^{\text{Tar}}}}$. Next, the CRB matrix of $\mathbf{p}_s$ can be derived as ${\mathrm{CRB}}_{\mathbf{p}_s^{\text{Tar}}}=\left[{\mathbf{J}}\left({\mathbf{p}_s^{\text{Tar}}}\right)\right]^{-1}$. Finally, CRB of coordinates ${x}_s, {y}_s$ can be derived as
\begin{equation}
\label{crb_derivation}
    {\mathrm{CRB}}_{x_{s},y_{s}} = {\mathrm{CRB}}_{x_{s}}+{\mathrm{CRB}}_{y_{s}}= \left[{\mathrm{CRB}}_{\mathbf{p}_s^{\text{Tar}}}\right]_{1,1}+\left[{\mathrm{CRB}}_{\mathbf{p}_s^{\text{Tar}}}\right]_{2,2}.
\end{equation}
\section{Proposed ARIS-Assisted ISAC System with DRL}
We aim to minimize the CRB of target coordinates by jointly optimizing the beamforming vectors $\{\mathbf{w}_{k,l}\}, \mathbf{R}_{s,l}$, the RIS phase shift $\mathbf{\Phi}_l$ and the ARIS's velocity $\mathbf{v}_l^\text{U}$ for all $l$. To this end, we formulate the optimization problem as
\begin{subequations}\label{p1}
\begin{align}
\textrm{(P1):} \ & \min_{\substack{\mathbf{v}^{\text{U}}_l,\{\mathbf{w}_{k,l}\},\mathbf{R}_{s,l}, \mathbf{\Phi}_l}} \,\, {\mathrm{CRB}}_{x_{s},y_{s}}\\
\text{s.t.}\quad & -W_{\max} \leq x^{\text{U}}_l,y^{\text{U}}_l \leq W_{\max}, \quad \forall l \in \mathcal{L}, \\
& v_{x,l}, v_{y,l} \leq v_{\max}, \quad \forall l \in \mathcal{L}, \\
& \Gamma_{k,l} \geq \Gamma_{th}, \forall k \in \mathcal{K}, \quad \forall l \in \mathcal{L}, \allowdisplaybreaks\\ 
& \sum\nolimits_{k\in \mathcal{K}}\Vert \mathbf{w}_{k,l}\Vert^{2}+\mathrm{t}\mathrm{r}\left[\mathbf{R}_{s,l}\right]\leq P_{\text{AP}},\,\, \forall l \in \mathcal{L}, \\ 
& \mathbf{R}_{s,l}\succeq 0 , \quad \forall l \in \mathcal{L},\\ 
& |[\boldsymbol{\Phi}_{l}]_{n,n}|=1, \quad \forall n \in \mathcal{N}, \forall l \in \mathcal{L}, 
\end{align}
\end{subequations}
where (\ref{p1}b) ensures that the ARIS travels within a $2W_{\max}$-side-length square; (\ref{p1}c) represents the constraints for the ARIS's maximum velocity; (\ref{p1}d) is the SINR constraint; (\ref{p1}e) and (\ref{p1}f) represent the constraints of transmit power and positive semi-definite condition of $\mathbf{R}_{s,l}$, respectively; (\ref{p1}g) guarantees that the reflection coefficient is a unit vector. 

To solve the non-convex problem (P1), we propose the DRL-based framework as in Algorithm 1. In the Markov decision process (MDP), the agent has a state $s_l$, and takes action $a_l$ in TS $l$. Given $s_l$ and $a_l$, as the agent proceeds with the various interactions in the environment, the agent obtains a reward $r_l$ and the next state $s_{l+1}$.

Among DRL techniques, the DDPG algorithm \cite{ddpg} has been actively adopted in dynamic environment since it is well-known to work well in the continuous action domain, such as ARIS trajectory. In the process of searching the optimal policy $\pi$ that maximizes the accumulated reward, the DDPG algorithm simultaneously learns the actor ($\mu$) and critic ($Q$) network with parameters $\theta^{\mu}$ and $\theta^{Q}$, respectively. The actor network learns the optimal policy that specifies the action for the state according to the following objective function according to the policy gradient method:
\begin{equation}
\label{actor}
    \nabla_{\theta^{\mu}}\mathcal{J} \approx \mathbb{E}\big[\nabla_{\theta^{\mu}} Q(s,a|\theta^Q)|_{s=s_l, a=\mu(s_l|\theta^{\mu})}\big],
\end{equation}
where $Q(s,a|\theta^Q)$ is the expected value of the state and action, and we call this value an action-value function. The critic network is a process of learning the action-value function, and instead of the existing greedy policy, it minimizes the loss function as follows,
\begin{equation}
\label{critic}
    \mathcal{L}(\theta^Q)=\mathbb{E}\big[\big(Q\big(s_l,a_l|\theta^Q\big)-y_l\big)^2\big],
\vspace{-2pt}\end{equation}
where sample $y_l=r_l+{\gamma}Q'\big(s_{l+1},\mu'(s_{l+1}|\theta^{\mu'})|\theta^{Q'}\big)$ and $\gamma \in [0,1]$ is the discount factor. Additionally, to increase learning convergence, the actor network and critic network have a copy of the target networks ($\mu', Q'$) to perform soft updates.

Here, to optimize the ARIS's trajectory based on the DDPG method, we define the state, action, and reward function in TS $l$ as follows:

\textbf{State: }Let $\mathcal{S}$ denote the system state space as $\mathcal{S}=\{s_l|s_l=\{x_l^\text{U},y_l^\text{U}, \tilde{x}_{s,l}, \tilde{y}_{s,l}\}$, where ${\tilde{x}}_{s,l},{\tilde{y}}_{s,l}$ are estimated $x,y$ coordinates of target in TS $l$, respectively. 

\textbf{Action: }Let $\mathcal{A}$ denote the system action space as $\mathcal{A}=\{a_l|a_l=\{v_{x,l}, v_{y,l}\}_{l\in \mathcal{L}}\}$.  

\textbf{Reward: }We define $r_l = 1/\mathrm{CRB}_{{\tilde{x}}_{s,l},{\tilde{y}}_{s,l}}-r_{p}$ as a reward function, where $\mathrm{CRB}_{{\tilde{x}}_{s,l},{\tilde{y}}_{s,l}}$ is the value obtained for the estimated target location, and $r_{p}$ is the penalty value when the ARIS is out of the given map. If ARIS moves off the map, it will be set to its previous location.
\begin{algorithm}[t]
	\caption{The proposed DRL-based method for ARIS-assisted ISAC systems}
	\hspace*{\algorithmicindent} \textbf{Input:} Structures of actor, critic and target network.
        \hspace*{\algorithmicindent} \textbf{Output:} $\mathbf{v}^{\text{U}}_l,\{\mathbf{w}_{k,l}\},\mathbf{R}_{s,l}, \mathbf{\Phi}_l$.
	\begin{algorithmic}[1]
    \State \textbf{Initialize: }Replay buffer $\mathcal{B}$, actor $\mu$, critic $Q$ and target network $\mu'$ and $Q'$ with $\theta^\mu$, $\theta^Q$, $\theta^{\mu'}\hspace{-4pt}\leftarrow\hspace{-1pt}\theta ^{\mu}$ and $\theta^{Q'}\hspace{-4pt}\leftarrow\hspace{-1pt}\theta^Q$;
    \For {Episode=1,.2,...,$E_{\max}$}
        \State Initialize environment setting;
    	\For {TS in $L_{\text{tot}}$}
    	    \State Execute action $a_l=\mu(s_l|\theta^\mu)+\mathcal{N}$; 
    	    \State Optimize phase shift $\mathbf{\Phi}_l$ and transmit beamformers \hspace*{\algorithmicindent}\hspace*{\algorithmicindent}$\mathbf{R}_{s,l}, \{\mathbf{w}_{k,l}\}$ in a changed environment by \cite{ruiizhang_RIS_ISAC};
         \State Obtain estimated target location $\tilde{x}_{s,l}, \tilde{y}_{s,l}$ via MLE;
          \State Obtain $r_l$ and $s_{l+1}$;
    	    \State Store transition $\big(s_l, a_l, r_l, s_{l+1}\big)$ in $\mathcal{B}$;
    	    
    	\EndFor
        \State Randomly sample a mini-batch transitions from $\mathcal{B}$;
    	    \State Update actor and critic network by (\ref{actor}) and (\ref{critic});
    	    \State Update target networks as $\theta^Q \leftarrow \tau \theta^Q+(1-\tau)\theta^{Q'}$
         \hspace*{\algorithmicindent}and $\theta^{\mu'} \leftarrow \tau \theta^{\mu}+(1-\tau)\theta^{\mu'}$;
    \EndFor
	\end{algorithmic} 
\end{algorithm}
The proposed method is provided in Algorithm 1 based on the settings above. After setting the initial weights and parameters, the proposed system model is operated for each episode. First, ARIS moves each time slot by adding a noise process $\mathcal{N}$ to the previous action obtained through the actor network for exploration. Next, we obtain the optimal solution for the remaining variables except for velocity variables $v_{x,l}, v_{y,l}$. The method of solving transmit beamforming and RIS phase shift through convex optimization \cite{ruiizhang_RIS_ISAC} can be applied to this system considering SINR constraints. Afterward, the estimated target coordinates $\tilde{x}_{s,l},\tilde{y}_{s,l}$ are determined through maximum likelihood estimation (MLE) based numerical methods \cite{crb}. Then, the CRB of the estimation target coordinates based on (\ref{crb_derivation}) can be derived from the optimized variable settings in TS $l$. When an episode ends, the networks are updated based on accumulated rewards and samples, and this process is repeated until the network converges.


\section{Simulation result}
\begin{table}[t]
    \scriptsize
    \vspace{-2pt}
    \caption{Simulation parameter setting}
    \label{parameter}
    \centering
    \vspace{-4pt}
    \begin{tabular}{cccccc} 
        \\[-1.8ex]\hline 
        \hline \\[-1.8ex] 
        \multicolumn{1}{c}{Parameter} & \multicolumn{1}{c}{Value} & \multicolumn{1}{c}{Parameter} & \multicolumn{1}{c}{Value}\\
        \hline \\[-1.8ex] 
        {$a$} & $3.5$                                & {$P_{\text{AP}}$} &  $40\;$dBm\\
        {$L_{\rm{tot}}$} & $12$                      & {$\gamma ^{\text {SI}}_{i,j}$} &  $-110\;$dBm \cite{clutter_user}\\
        {$W_{\max}$} & $100\;$m                      & {$\sigma^2_{s}$} &  $-110\;$dBm\\
        {$\Delta$} & $1\;$s                          & {$\sigma_k^2$} &  $-100\;$dBm \\
        {$H$} &  $50\;$m                             & {$\beta_0$} &  $-20\;$dB   \\ 
        {$B$} &  $1\;$MHz                            & {$\beta_s$} &  $-47\;$dB \cite{isac_fromthesky}   \\ 
        {$G_p$} &  $0.1B$ \cite{isac_fromthesky}     & {$\Gamma_{th}$} &  $10\;$dB  \\ 
        {$v_{\max}$} &  $8\;$m/s                     & {$\gamma$} &  $0.95$  \\ 
        {$M$} &  $16$                                & {$\tau$} &  $0.005$     \\ 
        {$N$} &  $16$                                \\
        \\[-1.8ex]\hline 
        \hline \\[-1.8ex] 
    \end{tabular}
    \vspace{-10pt}
\end{table} 
In this section, we provide several simulation results to validate the performance of the proposed method for ARIS-assisted ISAC systems. For benchmarks, we consider the following schemes:
\begin{itemize}
\item \textit{Fixed RIS}: The proposed algorithm optimizes the transmit beamforming and the ARIS's trajectory, while the RIS phase shift is fixed to the sensing target.
\item \textit{Without NSP}: All optimization variables are designed by the proposed algorithm without adopting the NSP-based receive beamforming scheme.
\end{itemize}
For simulations, we consider the parameter setting provided in Table \ref{parameter} by following \cite{ruiizhang_RIS_ISAC, isac_fromthesky}. We set $E_{\max}=500$ episodes with a learning rate $3 \times 10^{-4}$, the replay buffer capacity is 8000, the mini-batch size is $70$, and AdamOptimizer is used in the training stage. Both the actor and critic networks consist of fully connected hidden layers with $[300, 100, 100]$ neurons.
 \begin{figure}[t] 
\begin{center}
\includegraphics[width=1\columnwidth]{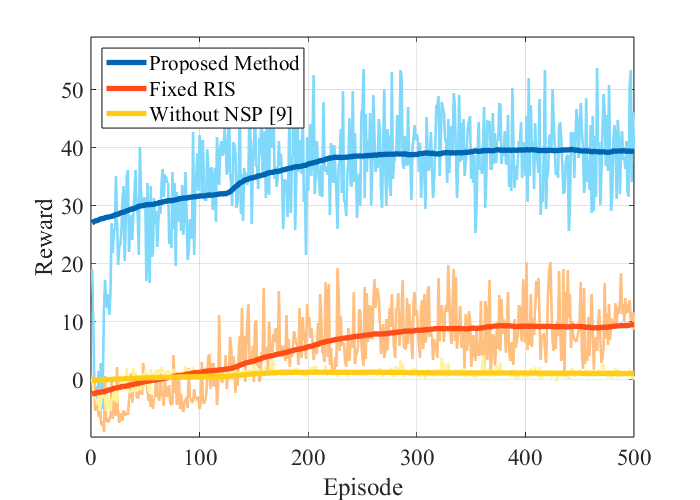}
\caption{Reward versus number of episodes.}
\label{reward}
\end{center}
\end{figure}

Fig. \ref{reward} illustrates the accumulated reward of the proposed DRL-based method according to the training episode. Initially, all schemes tend to have negative reward values due to the impact of penalties $r_p$ when ARIS goes off the map. While the Without NSP scheme achieves a reward close to $0$, the other schemes, including the proposed method, tend to converge starting from episode 200. In addition, the proposed method achieves the highest reward than other benchmark schemes. 
The sensing performance mentioned hereafter is analyzed based on the dataset obtained from the instance of the highest reward.

 \begin{figure}[t] 
\begin{center}
\includegraphics[width=1\columnwidth]{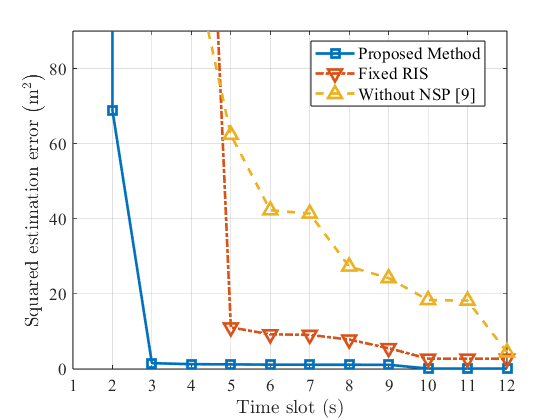}
\caption{The squared estimation error versus time slot. ($\Gamma_{th}=10$ dB).}
\label{crb}
\end{center}
\vspace{-15pt}
\end{figure}

 \begin{figure}[t]
  \centering
  \begin{tabular}{c@{\hspace{-13pt}} c }
   \hspace{-18pt} \includegraphics[width=.75\columnwidth, height = 0.65\columnwidth]{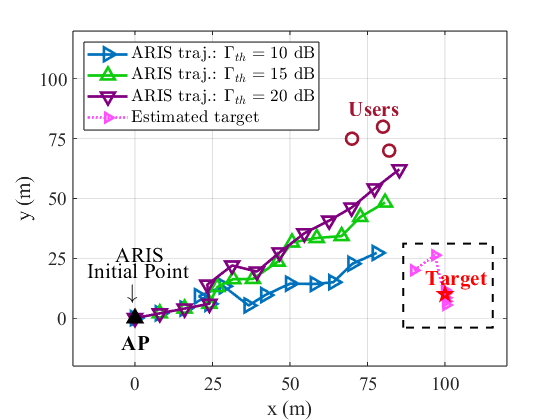} &
      \includegraphics[width=.35\columnwidth, height = 0.605\columnwidth]{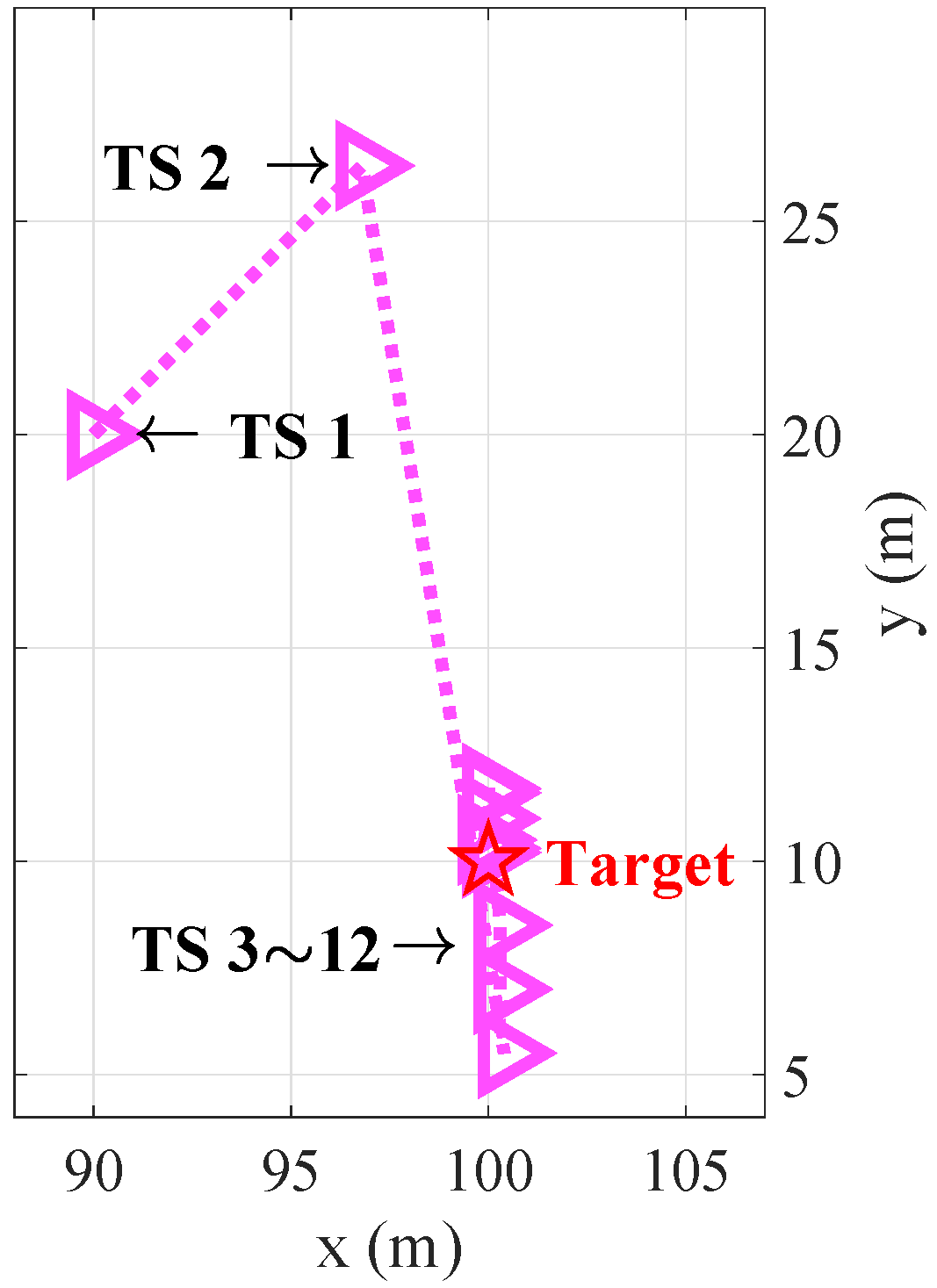} \\
        \small (a) &
      \small (b)
  \end{tabular}
  \caption{Optimal ARIS trajectories of the proposed method according to the different SINR threshold $\Gamma_{th}$. (The coordinates of the estimated target $\tilde{x}_{s,l}, \tilde{y}_{s,l}$ is depicted when $\Gamma_{th}=10$ dB.}
  \label{traj}
\end{figure}

In Fig. \ref{crb}, we compare the sensing performance of the proposed method with benchmark methods according to the time slot. Here, we use squared estimation error (SEE), which represents the error in the estimated target coordinates. The Without NSP scheme achieves the worst SEE because it does not suppress the effects of clutter echo and SI and thus obtains lower SNR values in (\ref{variance}). While the Fixed RIS method decreased SEE starting from TS $5$, the proposed method rapidly achieved the minimum SEE after TS $3$. It is verified that the proposed method, which jointly optimizes all variables and can cancel both SI and clutter echo, shows superior sensing performance and can rapidly stabilize battlefield situations. 

Fig. \ref{traj} depicts the optimized ARIS trajectories of the proposed method according to the different SINR threshold $\Gamma_{th}$. It can be observed that the SINR threshold determines the optimal location where ARIS will converge at the boundary between the target and the users. As the SINR threshold increases, there is a tendency to move towards the users rather than the target to accommodate more communication signals. In addition, we marked the estimated coordinates of the target (pink dashed line) when $\Gamma_{th}=10$ dB. As depicted in Fig. \ref{crb}, the proposed method can estimate the actual coordinates of the target almost similarly after TS $3$.
\section{Conclusion}
In this paper, we proposed a novel ARIS-assisted ISAC system for enhancing the awareness of practical battlefields. With DRL, we optimized the transmit beamforming at the access point, the ARIS's phase shift, and the ARIS's trajectory. In addition, we leverage a NSP-based receive beamforming scheme that suppresses self-interference and clutter echo signals. We validated that the proposed ISAC system significantly outperforms the conventional benchmark schemes in terms of sensing capability.
\appendix
\subsection{Derivation of measurement covariance $\mathbf{C}(\mathbf{d}_s)$ and FIM ${\mathbf{J}}\left({\mathbf{d}}_{s}\right)$}
First, we assume measurements $\widehat{{\mathbf{d}}}_{s}$ satisfies $\widehat{{\mathbf{d}}}_{s}\sim \mathcal{N} \left({\mathbf{d}}_{s},{\mathbf{C}}\left({\mathbf{d}}_{s}\right)\right)$ such that both the mean and covariance can depend on the distance parameter $\mathbf{d}_s$, where
$\widehat{{\mathbf{d}}}_{s}=\left[\widehat{d}_{s,1},...,\widehat{d}_{s,L_{\rm{tot}}}\right]^T$.
Based on the variance in (\ref{variance}), the measurement covariance $\mathbf{C}(\mathbf{d}_s)$ can be derived as
\begin{equation}
\begin{aligned}
    &{\mathbf{C}}\left({\mathbf{d}}_{s}\right)={\rm{diag}} \left([\sigma(d_{s,1})]^2, ..., [\sigma(d_{s,L_{\mathrm{tot}}})]^2 \right)\\&= \frac{a\sigma^2_{s}}{P_{\text{AP}}G_{\rm{p}}\beta_s}{\rm{diag}}\left(\frac{\left[d_{s,1} \right]^4}{\lVert\mathbf{Z}_{1}^\text{Echo}\rVert^2},...,\frac{\left[d_{s,{L_{\mathrm{tot}}}} \right]^4}{\lVert\mathbf{Z}_{L_\text{tot}}^\text{Echo}\rVert^2}\right),
\end{aligned}
\end{equation}
where $\mathbf{Z}_l^\text{Echo}=(\mathbf{G}_l^{\text{AP,U}})^T\mathbf{\Phi}_{l}^T\mathbf{H}_l^\text{Tar}\mathbf{\Phi}_l\mathbf{G}_l^{\text{AP,U}}$. By following \cite{measurementcovariance}, the general expression of $p,q$-th element in FIM can be expressed as
\begin{equation}
\begin{aligned}
&\left[{\mathbf{J}}\left({\mathbf{d}}_{s}\right)\right]_{p,q}=\left[\frac{\partial {\mathbf{d}}_s}{\partial d_{s,p}}\right]^T{\bf{C}}^{-1}\left({\mathbf{d}}_s\right)\left[\frac{\partial {\mathbf{d}}_s}{\partial d_{s,q}}\right]\\&+\frac{1}{2}{\rm{tr}}\left[{\bf{C}}^{-1}\left({\mathbf{d}}_s\right)\frac{\partial{\bf{C}}\left({\mathbf{d}}_s\right)}{\partial d_{s,p}}{\bf{C}}^{-1}\left({\mathbf{d}}_s\right)\frac{\partial{\bf{C}}\left({\mathbf{d}}_s\right)}{\partial d_{s,q}}\right].
\end{aligned}
\end{equation}
Due to the diagonal nature of the measurement covariance, each component of FIM becomes 0, if $p \neq q \in \{1,...,L_\text{tot}\}$. To this end, the closed form of ${\bf{J}}\left({\mathbf{d}}_s\right)$ is 
\begin{equation}
\begin{split}
&{\bf{J}}\left({\mathbf{d}}_s\right)\\&= \text{diag} \left(
\frac{1}{\left[\sigma(d_{s,1})\right]^2} +  \frac{8}{\left[d_{s,1} \right]^2},..., 
\frac{1}{\left[\sigma(d_{s,L_{\text{tot}}})\right]^2} + \frac{8}{\left[d_{s,L_{\text{tot}}} \right]^2}
\right).
\end{split}
\end{equation}

\bibliographystyle{IEEEtran}
\bibliography{ref}

\begin{thebibliography}{10}
\providecommand{\url}[1]{#1}
\csname url@samestyle\endcsname
\providecommand{\newblock}{\relax}
\providecommand{\bibinfo}[2]{#2}
\providecommand{\BIBentrySTDinterwordspacing}{\spaceskip=0pt\relax}
\providecommand{\BIBentryALTinterwordstretchfactor}{4}
\providecommand{\BIBentryALTinterwordspacing}{\spaceskip=\fontdimen2\font plus
\BIBentryALTinterwordstretchfactor\fontdimen3\font minus \fontdimen4\font\relax}
\providecommand{\BIBforeignlanguage}[2]{{%
\expandafter\ifx\csname l@#1\endcsname\relax
\typeout{** WARNING: IEEEtran.bst: No hyphenation pattern has been}%
\typeout{** loaded for the language `#1'. Using the pattern for}%
\typeout{** the default language instead.}%
\else
\language=\csname l@#1\endcsname
\fi
#2}}
\providecommand{\BIBdecl}{\relax}
\BIBdecl

\bibitem{isac_overview}
F.~Liu, Y.~Cui, C.~Masouros, J.~Xu, T.~X. Han, Y.~C. Eldar, and S.~Buzzi, ``Integrated sensing and communications: Toward dual-functional wireless networks for 6{G} and beyond,'' \emph{IEEE J. Sel. Areas in Commun.}, vol.~40, no.~6, pp. 1728--1767, 2022.

\bibitem{military}
C.~B. Barneto, S.~D. Liyanaarachchi, M.~Heino, T.~Riihonen, and M.~Valkama, ``Full duplex radio/radar technology: The enabler for advanced joint communication and sensing,'' \emph{IEEE Wireless Commun.}, vol.~28, no.~1, pp. 82--88, 2021.

\bibitem{clutter_user}
Z.~He, W.~Xu, H.~Shen, D.~W.~K. Ng, Y.~C. Eldar, and X.~You, ``Full-duplex communication for {ISAC}: Joint beamforming and power optimization,'' \emph{IEEE J. Sel. Areas Commun.}, vol.~41, no.~9, pp. 2920--2936, 2023.

\bibitem{ris}
Q.~Wu and R.~Zhang, ``Intelligent reflecting surface enhanced wireless network via joint active and passive beamforming,'' \emph{IEEE Trans. Wireless Commun.}, vol.~18, no.~11, pp. 5394--5409, 2019.

\bibitem{ruiizhang_RIS_ISAC}
X.~Song, X.~Qin, J.~Xu, and R.~Zhang, ``Cramér-rao bound minimization for {IRS}-enabled multiuser integrated sensing and communications,'' \emph{IEEE Trans. Wireless Commun.}, pp. 1--1, 2024.

\bibitem{ARIS}
H.~Lu, Y.~Zeng, S.~Jin, and R.~Zhang, ``Aerial intelligent reflecting surface: Joint placement and passive beamforming design with 3{D} beam flattening,'' \emph{IEEE Trans. Wireless Commun.}, vol.~20, no.~7, pp. 4128--4143, 2021.

\bibitem{ARIS_ISAC}
J.~Xu, D.~Li, Z.~Zhu, Z.~Yang, N.~Zhao, and D.~Niyato, ``Anti-jamming design for integrated sensing and communication via aerial {IRS},'' \emph{IEEE Trans. Commun.}, pp. 1--1, 2024.

\bibitem{ddpg}
\BIBentryALTinterwordspacing
T.~P. Lillicrap \emph{et~al.}, ``Continuous control with deep reinforcement learning,'' 2015. [Online]. Available: \url{arXiv:1509.02971}
\BIBentrySTDinterwordspacing

\bibitem{benchmark}
J.~Luo, T.~Liang, C.~Chen, and T.~Zhang, ``A {UAV} mounted {RIS} aided communication and localization integration system for ground vehicles,'' in \emph{Proc. 2022 IEEE Int. Conf. Commun. (ICC) Workshops}, May 2022, pp. 139--144.

\bibitem{self_interference_not_neglect}
J.~S. Yeom, Y.-B. Kim, and B.~C. Jung, ``Spectrally efficient uplink cooperative {NOMA} with joint decoding for relay-assisted {I}o{T} networks,'' \emph{IEEE Internet Things. J.}, vol.~10, no.~1, pp. 210--223, 2023.

\bibitem{crb}
S.~M. Kay, \emph{Fundamentals of statistical signal processing: {E}stimation theory}.\hskip 1em plus 0.5em minus 0.4em\relax Prentice-Hall, Inc., 1993.

\bibitem{nsp}
C.~B. Barneto, S.~D. Liyanaarachchi, T.~Riihonen, M.~Heino, L.~Anttila, and M.~Valkama, ``Beamforming and waveform optimization for {OFDM}-based joint communications and sensing at mm-waves,'' in \emph{Proc. Asilomar Conf. on Signals, Syst., Comput.}, Nov. 2020, pp. 895--899.

\bibitem{RIS_ISAC}
R.~Liu, M.~Li, H.~Luo, Q.~Liu, and A.~L. Swindlehurst, ``Integrated sensing and communication with reconfigurable intelligent surfaces: Opportunities, applications, and future directions,'' \emph{IEEE Wireless Commun.}, vol.~30, no.~1, pp. 50--57, 2023.

\bibitem{delta_seconds}
S.~Yoo, S.~Jeong, J.~Kim, and J.~Kang, ``Cache-assisted mobile edge computing over space-air-ground integrated networks for extended reality applications,'' \emph{IEEE Internet Things. J.}, vol.~11, no.~10, pp. 18\,306--18\,319, 2024.

\bibitem{uav_channel}
Q.~Wu, Y.~Zeng, and R.~Zhang, ``Joint trajectory and communication design for multi-{UAV} enabled wireless networks,'' \emph{IEEE Trans. Wireless Commun.}, vol.~17, no.~3, pp. 2109--2121, 2018.

\bibitem{self-interference}
M.~Temiz, E.~Alsusa, and M.~W. Baidas, ``A dual-function massive {MIMO} uplink {OFDM} communication and radar architecture,'' \emph{IEEE Trans. Cogn. Commun. Netw.}, vol.~8, no.~2, pp. 750--762, 2022.

\bibitem{isac_fromthesky}
X.~Jing, F.~Liu, C.~Masouros, and Y.~Zeng, ``{ISAC} from the sky: {UAV} trajectory design for joint communication and target localization,'' \emph{IEEE Trans. Wireless Commun.}, pp. 1--1, 2024.

\bibitem{uav_ISAC_crb}
J.~Wu, W.~Yuan, and L.~Bai, ``On the interplay between sensing and communications for {UAV} trajectory design,'' \emph{IEEE Internet Things J.}, vol.~10, no.~23, pp. 20\,383--20\,395, 2023.

\bibitem{snr_inverse_proportional}
F.~Liu, W.~Yuan, C.~Masouros, and J.~Yuan, ``Radar-assisted predictive beamforming for vehicular links: Communication served by sensing,'' \emph{IEEE Trans. Wireless Commun.}, vol.~19, no.~11, pp. 7704--7719, 2020.

\bibitem{measurementcovariance}
S.~Kay, \emph{Fundamentals Of Statistical Processing, Volume II: Detection Theory}, ser. Prentice-Hall signal processing series.\hskip 1em plus 0.5em minus 0.4em\relax Pearson Education, 2009.

\end{thebibliography}

\end{document}